\documentclass{mystylePoS}

\usepackage{amsmath}

\def\b{{\bf b}}

\def\z{{\bf z}}
\def\q{{\bf q}}

\def\B{{\cal B}}

\def\Z{{\cal Z}}

\def\p+{\! + \!}
\def\m-{\! - \!}
\def\r={\! = \!}
\def\Im{{\mathrm{Im}\,}}
\def\Re{{\mathrm{Re}\,}}

\def\scss{\scriptscriptstyle}
\def\N{{\scss N}}

\def\phid{\phi^\dagger}
\def\r3P{r_{{\scss 3P}}}

\title{Reaction-diffusion approach in soft diffraction}

\ShortTitle{Reaction-diffusion approach in soft diffraction}

\author{{Rodion Kolevatov}\thanks{Also at Saint-Petersburg State University, Department of High-energy physics, 
  Ulyanovskaya 1, 198504, Saint-Petersburg, Russia. }\\
        SUBATECH, Ecole des Mines de Nantes, 4 rue Alfred Kastler, 44307 Nantes Cedex 3, France \\
        E-mail: \email{rodion.kolevatov@fys.uio.no}}

\author{Konstantin Boreskov\\
        {Institute of Theoretical and Experimental Physics, 117259, Moscow, Russia}\\
        E-mail: \email{boreskov@itep.ru}}

\abstract{We apply the reaction-diffusion (stochastic) approach to the numerical calculation of the elastic amplitude in
the Reggeon Field Theory (RFT) and its single diffractive cut. Fits to the total, integrated and differential elastic cross sections with account of all Pomeron loops are reported together with all-loop calculation of the single difraction dissociation cross section.}

\FullConference{Prepared for the proceedings of the XXI International Workshop\\ High Energy Physics and Quantum Field Theory,\\
		June 23 -- June 30, 2013 \\
		Saint Petersburg Area, Russia}

\begin{document}

\section{Elastic and inelastic diffraction.}
A substantial part of the total interaction cross section of hadrons at high energies is du to elastic and inelastic diffractive interactions. Presense of elastic scatttering even at highest interaction energies  is dictated by the unitarity of the elastic scattering amplitude, while the inelastic diffraction dissociation represents a special class of events. In these events either one or both of the interacting hadrons dissociate into a hadron system (reffered to as single (SD) and double (DD) diffractive dissociation respectively) or both of the hadrons remain intact with particle production only at midrapidity. The latter class of events is refferred to as central diffraction (CD). 

An elegant interpretation of inelastic SD and DD interactions  for the case of a low invariant mass of the diffracitvely produced system $M_X\sim m_{\rm hadron}$ has been given by Good and Walker \cite{GoodWalker}. In their formalism the incoming hadron is represented as a superposition of the eigenstates of the scattering operator. The coherence of the superposition is broken upon scattering under the condition that not all of the eigenamplitudes equal each other. In particular in the black disk limit, when the eigenamplitudes equal unity in the impact parameter representation, the inelastic diffraction exists at the edge of the disk only and is asymptotically suppressed at high energies compared to the total inelastic cross section. The Good--Walker formalism is used in a number of models \cite{Ryskin:2011qe,Gotsman:2008tr,Ostapchenko:2010gt}, including the one of the authors to fit the data on low-mass diffraction.

On the other hand a good description of the data on low-mass diffraction at various center of mass energies has been achieved in the OPER model \cite{Boreskov:1972mc}. In this approach the incident nucleon in the diffraction dissociation event fluctuates into a $\pi$-meson--nucleon pair prior to the interaction with the subsequent scattering of either particle of the pair on the target. Applied to Good--Walker formalism with the minimal choice of two scattering eigenstates (two channes) this implies one of the channels to have a significantly larger coupling and interaction radius than the other. For the diffractive production of states with large invariant mass $M_X \gg m_{\rm hadron}$ the Good--Walker formalism is no longer applicable due to its explicit separation of diffractive and multiparticle states. 

For $M_X\gg m_{\rm hadron}$ the differential SD cross section $\frac{d\sigma_{SD}}{dt dM_X^2}$ at fixed $t$ excibits a characteristic $\frac{1}{M_X^2}$ behaviour  in contrast with the resonance-like structure at low masses, $M_X\sim m_{\rm hadron}$ \cite{Goulianos:1982vk}. This behaviour comes out naturally in the Reggeon Field Theory (RFT) based models. The elastic scattering amplitude in the RFT is given by the exchanges of Pomerons with vacuum quantum number in the $t$-channel. The cuts of the elastic scattering amplitude give the cross section for various inelastic processes. In particular events with rapidity gaps come form the cut graphs with Pomeron interactions (enhanced) and loops when the cut goes in between the Pomerons.

At the same time an increase with energy of the total interaction cross section indicates that the intercept of the Pomeron trajectory is larger than unity. This implies power-like growth of the total cross section ($\sigma_{tot}^{1\mathbb P} \sim s^{\Delta}$, $\Delta>0$) which is in contradiction with the unitarity constraint ($\sigma_{tot}\le C \ln^2 s$). The enhanced and loop contributions together with multipomeron exchahges are essential for taming the growth and restoring the unitarity. The Reggeon Field Theory, a systematic way for accounting of these graphs, has been formulated it the works of Gribov \cite{GribovRFT}.

\section{The reaction-diffusion approach in Reggeon Field Theory}

The elastic scattering amplitude in the RFT is obtained as a convolution of the process-dependent vertices and the process-independent Green functions which contains all the dynamics of the interaction. The Green function are obtained withtin the 2+1 dimentional field theory with the Lagrangian
\begin{equation}
 \mathcal{L} = \frac{1}{2} \phi^\dagger (\overleftarrow{\partial_y} - \overrightarrow{\partial_y})\phi
 - \alpha' (\nabla_\b \phid)(\nabla_\b \phi)   + \Delta \phid\phi +\mathcal{L}_{int}.
\label{eq:LRFT}
\end{equation}
A minimal choice for $\mathcal{L}_{int}$ which is dictated by the presence of inelastic high-mass diffraction is the triple Pomeron vertex. However from the phenomenological point of view the more compicated vetices are not excluded and a number of models has been suggested which involve infinite sets of $m$Pomeron$\to n$Pomeron couplings \cite{Ryskin:2011qe,Gotsman:2008tr,Ostapchenko:2010gt}. 

The reaction-diffusion (RD) or stochastic approach used by the authors implies the ``almost minimal'' choice for the interaction $\mathcal{L}_{int}$. It was observed \cite{Grassberger:1978pr} that a system of classical particles (``partons'') on two dimentional plane with certain evolution rules admits a field-theoretical description with the Lagrangian of the RFT with interaction term containing Pomeron scattering in addition to  the triple coupling:
\begin{equation}
\mathcal{L}_{int} = i\,\r3P \phi^\dagger \phi (\phi^\dagger + \phi) + \chi {\phi^\dagger}^2 \phi^2  \label{eq:Lint}
\end{equation}

Partons af the stochastic system are allowed to move chaotically (characterized by diffusion coefficient $D$), split, $A\rightarrow A+A$, with probability per unit time $\lambda$, or die, $A\rightarrow \emptyset$, with a death probability $m_1$. When two partons are brought within the reaction range $a$ due to the diffusion, they can pairwise fuse, $A+A\rightarrow A$, or annihilate, $A+A\rightarrow \emptyset$ with the rates $\nu$ and $m_2$ correspondingly. The stochastic system of partons can be described by the symmetrized probability densities
$\rho_\N(y;\B_N)$  with normalization  $
\sum_N \frac{1}{N!}\int d\B_\N \rho_\N (y;\B_\N) = \sum_M p_\N (y) =1
$  (here $\Z_s\equiv\{\z_1,\ldots, \z_s\}$). An equivalent description in terms of inclusive $s$-parton distributions
\begin{equation}
 f_s(y;\Z_s)= \sum_{N\ge s} \cfrac{1}{(N-s)!} \int\! d\B_N \;\rho_\N(y;\B_N)
 \prod_{i=1}^{s} \delta(\z_i-\b_i), \label{def_f}
\end{equation} 
allows to establish connection with the Reggeon Field Theory: the set of evolution equations for the $f_s(y;\Z_s)$ coincides with the evolution equations for  the \emph{exact} Green functions of the RFT with the interaction Lagrangian \eqref{eq:Lint}.

Phenomenological parameters of the Lagrangian \eqref{eq:LRFT} have direct correspondence with the rates of the stochastic system (see table 1). The parton interaction distance $a$ serves as a regularization parameter for the Pomeron loops. For given values of the coupling $r_{3P}$ and the scale  $\epsilon \equiv \pi a^2$ the quartic coupling $\chi$ can be varied. % which thus has a physical meaning of the size of a Pomeron. This correspondence is outlined in table \ref{TAB:RFT-ST-rel} where $\epsilon \equiv \pi a^2$.
\begin{table}[h]
\centering
\caption{Relation between the parameters of the RFT and those of the stochastic approach.}
\begin{tabular}{|c||c|c|c|c|c|c|}
 \hline
 RFT & $\alpha'$ & $\Delta$ & $\r3P$, $\mathbb P$ splitting vertex& $\r3P$, $\mathbb P$ fusion vertex & $\chi$, $2{\mathbb P}\to 2{\mathbb P}$\\
 \hline
 RD-approach & $D$ & $\lambda - m_1$ & $\lambda \sqrt{\epsilon}$ & $({m_2}+\tfrac{1}{2}{\nu})\sqrt{\epsilon}$ &  $\tfrac{1}{2} ({m_2}+{\nu})\epsilon $\\
 \hline
\end{tabular}
\label{TAB:RFT-ST-rel}
\end{table}
\vskip-1mm
This correspondence allows to obtain numerically various quantities in the Reggeon Field theory with account of all loops following a Monte-Carlo evolution of the RD system. 

The procedure for computing the elastic scattering amplitude and its single diffractive cut was described in \cite{Boreskov:2001nw,Kolevatov:2011nf,Kolevatov:2012vu}. The amplitude is given by the convolution at some linkage point $y$ in rapidity of the the projectile- ($f_s$) and target associated ($\tilde f_s$) inclusive distributions %(grey blocks in fig. 1{\it a}) 
according to the general rules of the Reggeon field theory:
\begin{equation}
 T^{\rm el}({\bf b}, Y) %= \langle A | T | \tilde{A} \rangle 
 =\sum_{s=1}^\infty \frac{(-1)^{s-1}} {s!} \int d\Z_s
 d\tilde \Z_s f_s(y;\Z_s ) \tilde f_s(Y-y;\tilde \Z_s) \prod_{i=1}^s g ({\bf z}_i - \tilde {\bf z}_i - {\bf b}) .
\label{TST}
\end{equation} 
\vskip-2mm
\noindent Here $g$ are some narrow functions normalized to $\int g({\bf b}) d^2{\bf b}=\epsilon$. The most efficient way to compute this convolution is to do it on event by event basis with setting the linkage point to the target rapidity and doing subsequent Monte-Carlo average. 

The inclusive $s$-parton distributions at projectile and target rapidities coincide with the hadron--$s$-Pomeron vertices \cite{Kolevatov:2011nf}:
\vskip-4mm
\begin{equation}
 f_s(y=0;\Z_s)  %\propto
\equiv \mu_s p_s(\Z_s)=  \epsilon^{s/2} \mathcal{N}^{(s)}(\Z_s). \label{fs-vertex}
\end{equation}
In particular, the two-channel eikonal vertices correspond to the superposition of two Poissonian distributions for the number of partons at zero evolution time. 
%This defines the distribution of partons at zero evolution time in number and positions in the transverse plane. In accordance with it the initial configurations of partons should be generated within the first step of the Monte-Carlo averaging procedure. 
Upon the Monte-Carlo evolution of initial random parton configuration one gets a set of $N$ partons at certain positions $\hat b_i$ in the transverse plane. The event realization of inclusive distribution is thus  
$ f^{event}_s (\B_s)= \sum_{\{i_1\ldots i_{s}\}\in \{ 1\ldots N\}} \delta(b_1-\hat b_{i_1})\ldots \delta (b_{s} - \hat b_{i_{s}})$ and upon convolution with the set of target--$n$-Pomeron vertexes leads to: \vskip-4mm
\begin{equation}
T_{\rm sample}^{\rm el} ({\bf b}) = \sum_{s=1}^{N} (-1)^{s-1} \tilde \mu_s \epsilon^s \sum_{i_1<i_2 \ldots <
i_s} \tilde p_s(\hat {\bf x}_{i_1} - {\bf b}, \ldots ,\hat {\bf x}_{i_s}-{\bf b}). \label{T-onesample}
\end{equation} \nopagebreak The actual value of the elastic amplitude as a function of the impact parameter $T_{el}(b,Y=\ln s)$ is computed by making Monte-Carlo average of \eqref{T-onesample}. The numerical procedure was described in detail in \cite{Kolevatov:2011nf}.

Using the Lagrangian \eqref{eq:LRFT} implies that the elastic scattering amplitude we get is purely imaginary, $A_{P}(b) = iT_{el}(b)$, which is satisfied only approximately. At lower energies ($\sqrt s \lesssim 100$~GeV) we add contributions from two secondary trajectories with positive and negative signature to improve the quality of data description and assume that the real part is dominated by these contributions. For the elastic $pp$/$p\bar p$ scattering amplitude this gives:
\begin{eqnarray}
%\text{$pp$:}
& & \left.\Im f_{pp/p\bar p}(b)\right|_{\sqrt s \lesssim 100 {\rm GeV}} = \Im A_P(b)+ \left[\Im A_{R_+}(b)\pm \Im A_{R_-}(b)\right]\left[\vphantom{\Im A_{R_+}(b)} 1-\Im A_P(b) \right] \label{amplppsec}\\
                 & &\left. \Re f_{pp/p\bar p}(b) \right|_{\sqrt s \lesssim 100 {\rm GeV}}= \left[\Re A_{R_+}(b)\pm \Re A_{R_-}(b)\right]\left[\vphantom{\Re A_{R_+}(b)}1 - \Im A_P(b) \right] \nonumber
\end{eqnarray}
with $\displaystyle A_{R_\pm} (y, b) = \eta_\pm \beta_\pm^2 \frac{\exp(\Delta_\pm y)}{2 \alpha'_\pm y + 2 R_\pm^2} \exp\left( -\frac{b^2}{4(\alpha'_\pm y + R^2_\pm)} \right)
$  and
$ \displaystyle
\eta_\pm  =  \pm i - \frac{1 \pm \cos [\pi (\Delta_\pm+1)]}{ \sin [\pi (\Delta_\pm + 1)]}
$. We use Gaussian parameterization of Reggeon-hadron vertices.

For the energies of UA4 and higher contribution of secondary trajectories to the amplitude is negligible. Here we evaluate the real part of the amplitude from the Gribov--Migdal relation \cite{Gribov-Migdal}:
\begin{equation}
 \eta \equiv \left.\frac{\Re M(s,t)}{\Im M(s,t)}\right|_{t=0} \simeq \frac{\pi}{2} \frac{1}{\Im M(s,t=0)} \frac{d\Im M(s,t)}{d\ln s}. \label{eq:GrMi}
\end{equation}
where $ M(s, t= -{\q}^2) = \int d^2q\, e^{-i\q\b} f(Y=\ln s,\b)$ is the amplitude in the transverse momentum representation. Though \eqref{eq:GrMi} relates real and imaginary part of the forward scattering amplitude, that is, amplitude $f(b)$ integrated over the impact parameter, we make use of it extrapolating to arbitrary values of $b$. 

Once the ampitude $f(b)$ is computed, the total and elastic cross sections are expressed as: %in the usual way:
%\vspace{-2mm}
\begin{eqnarray}
&& \sigma^{\rm tot} (Y) = 2\!\int d^2b\, \Im f(Y,{\bf b}) ~,\\
&& \sigma^{\rm el} = \int d^2b\, |f(Y,{\bf b})|^2 ,\\
&& \left.\frac{d\sigma^{\rm el}}{dt}\right|_{t=-q^2} = \frac{1}{4\pi}\, \left|M(Y,{\bf q})\right|^2 =\pi \left| \int f(Y, b) J_0(qb) b db \right|^2.
\end{eqnarray}

\setlength{\textfloatsep}{15pt plus 2pt minus 4pt}
\setlength{\floatsep}{15pt plus 2pt minus 4pt}

%\begin{figure}[htbp]
% \centering
% \includegraphics[width=0.8\hsize]{figs/el+SD.pdf}
% \caption{Diagrams taken into account in calculation of elastic amplitude ({\it a--c}) and its SD cut ({\it d--g}).}
%\end{figure}

A single diffractive cut of the amplitude with a requirement of separation betwen the elastic scattered hadron the diffractively produced system to be at least $y_{\rm gap}$ rapidity units (with elastic contribution included) can be computed as a sum of two terms:% \footnote{We are currently preparing a paper with a complete derivation of eq. \eqref{eq:SDcut}.}:%\cite{st-new}:
\vspace{-1mm}
\begin{equation}
T^{\rm SD cut} ({\bf b},Y,y_{\rm gap})= 2 T^{\rm el}({\bf b},Y) - T'({\bf b},Y,y_{\rm gap}). \label{eq:SDcut}
\end{equation}
\vspace{-1mm}
The term $T'({\bf b},Y,y_{\rm gap})$ is computed in the same way as the elastic amplitude by making a Monte-Carlo average of \eqref{T-onesample} with only the distinction in preparation of the projectile-associated set of partons. The evolution starts with \emph{two sets} which evolve \emph{independently} up to the evolution time $y_{\rm gap}$ corresponding to the width of the rapidity gap. At that point the resulting partons are \emph{combined into a single set} which further evolves in the standard way from $y_{\rm gap}$ up to the target rapidity $Y$. Thus both, the elastic scattering amplitude and its single diffractive cut are computed within \emph{the same numerical framework.} Varying size of the gap $y_{\rm gap}$ it is possible to compute the differential diffractive cross section $\frac{d\sigma_{SD}}{dM_X^2}$.

\section{Parameters and results}

Prior to doing all-loop calculation we perform a two-channel eikonal fit to total and elastic cross sections, trying at the same time to obtain a good description of differential elastic cross sections and requiring the low-$M_X^2$ diffractive cross section to be $\sigma_{SD}\approx 1.5$~mbn at $\sqrt s = 35GeV/c$ in accord with ISR data \cite{Webb:1974nb,deKerret:1976ze}.  We use two-channel eikonal hadron--Pomeron vertices, $ \displaystyle 
 \mathcal{N}^{(s)}(\Z_s) = C_1 \beta_1^s\prod p_1({\bf z}_i) +C_2 \beta_2^s\prod p_2({\bf z}_i)$ %\label{all-loop-vert}
with gaussian profile
$
p_{1/2}({\bf z}_i)=\frac{1}{2 \pi R_{1/2}^2} \exp{\left(-\frac{{\bf z}_i^2}{2 R_{1/2}^2}\right)} %\label{cprof}
$
and parameterize for convenience $ \beta_{1/2}=\beta_P (1\pm \eta)$.% \label{defeta}
The lowest order contrbution of secondary trajectories and the real part of the amplitude are also accounted as described above.
 From the eikonal fit we fix the values of $C_1=0.1, C_2=1-C_1=0.9$ and $\eta=0.65$ and use them in the all loop calculation.  The values for other paremeters serve as the input for the first step of the all-loop fit. These values except of the intercept, are only slightly modified when doing the all-loop fit.

\begin{floatingtable}{
\begin{tabular}{|l|ccc|}
 \hline Trajectory & $\mathbb{P}$ & $R_+$ & $R_-$ \\
\hline  $\Delta_{P/+/-}$ &  0.19 & -0.27  &   -0.55 \\ 
  $\alpha'$, GeV$^{-2}$ &  0.258 & 0.70 &  1.0\\
  $R^2$, GeV$^{-2}$ & 7.79/0.26  & 3.0 &   9.5 \\
  $\beta_{P/+/-}$, GeV$^{-1}$ & 9.4 & 8.0 &   3.3 \\
  \hline
\end{tabular}}
%\noindent \footnotesize {\bf TABLE 2.} Parameters of the trajectories and their couplings to proton.
%\vskip-3mm
\caption{Fitted parameters for the trajectories.}
\end{floatingtable}  

For the all-loop calculation in addition we fix apriori the regularization scale $a=0.036$~fm$=0.182$~GeV$^{-1}$, fix the triple coupling value at \linebreak ${r_{3P}=0.087}$~GeV$^{-1}$ according to \cite{Kaidalov:1979jz}\footnote{We use a different normalization of $r_{3{\mathbb P}}$ (see \cite{Kolevatov:2011nf}).} and fix $2\to 2$ coupling $\chi$ by setting $\nu=1/2\lambda$ and $m_2=0$ (see tab.~1).  This is the same choice as for ``set 3'' in \cite{Kolevatov:2012vu} where dependence on the scale $a$ and coupling $\chi$ was also studied.

% gives $\chi =2.87\times 10^{-2}$~GeV$^{-2}$.
\pagebreak 

The preliminary result for the all-loop fit to the total and elastic cross sections are plotted in fig. 2 and the values of the fitted parameters are outlined in tab.~2. For comparison we plot also the differential elastic cross section for ``set 3'' of \cite{Kolevatov:2012vu}. We observe that data for differential elastic cross section $d\sigma_{el}/dt$ at larger $|t|$ favor the configuration when couplings $\beta_{1/2}$  and radii $R_{1/2}$ for the two channels significantly differ with much lower probability for the channel with larger coupling. This is in accord with the expectations from the OPER model \cite{Boreskov:1972mc}.

\begin{figure}[!h]
 %\centering 
 \hspace{-6mm}\hbox{\includegraphics[width=0.4\hsize]{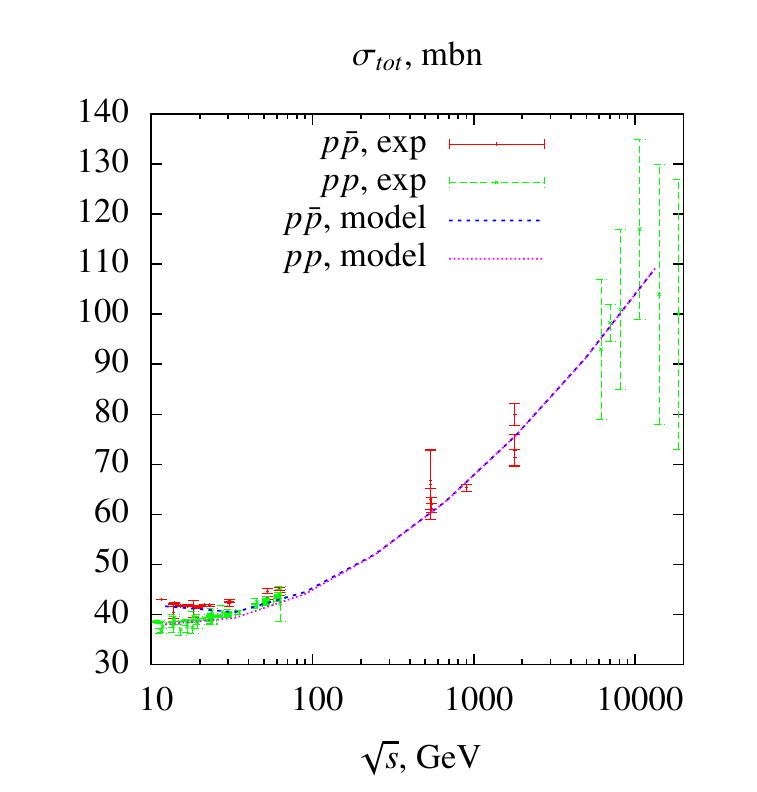}\hspace{-8mm}\includegraphics[width=0.4\hsize]{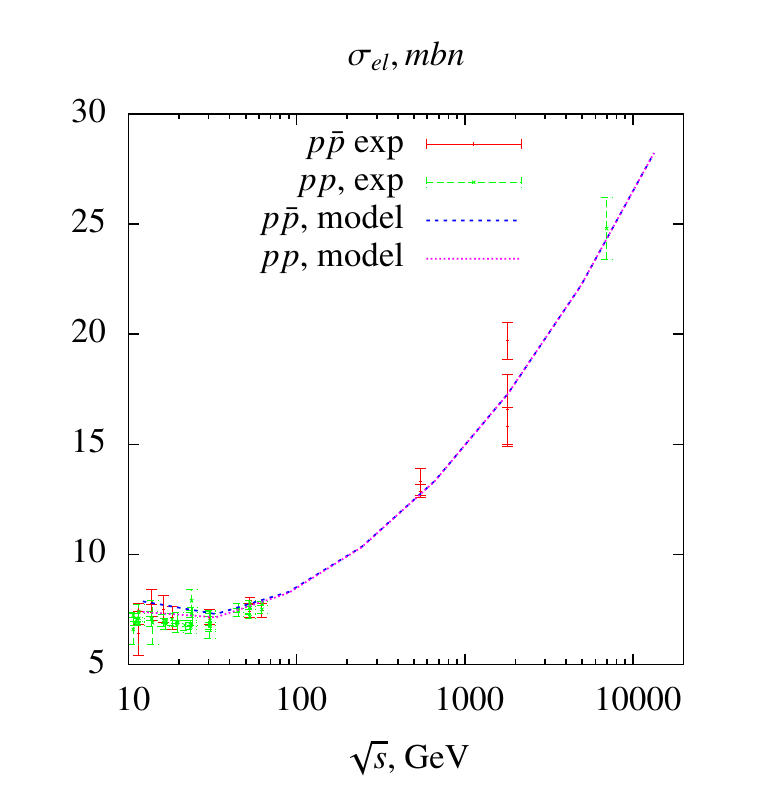}\hspace{-10mm}\includegraphics[width=0.4\hsize]{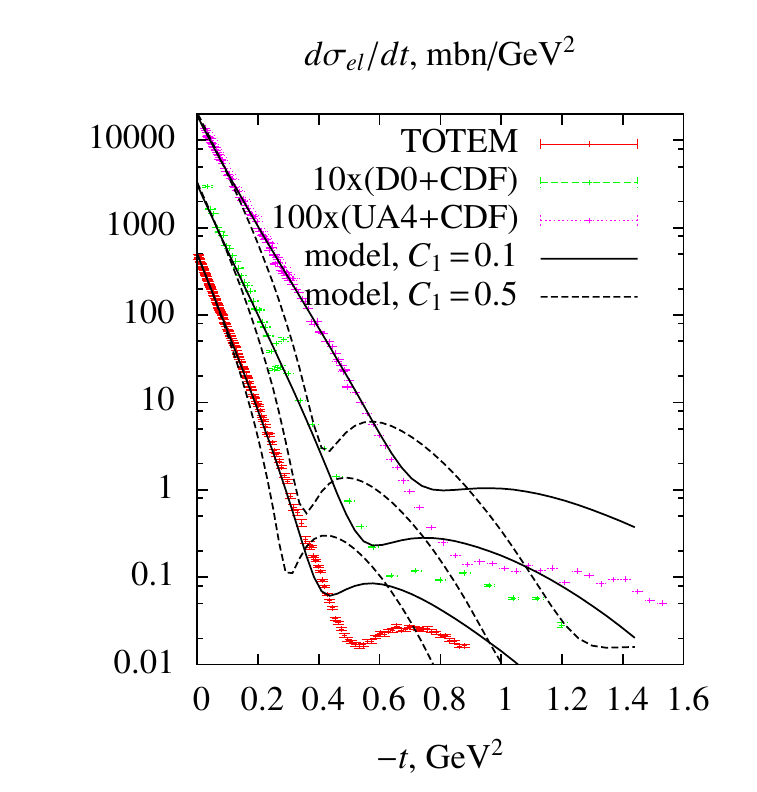}}
 \caption{Left, center: all-loop fits to total and elastic cross sections. Right: differential elastic cross section at $\sqrt s =546,1900$ and 7000 GeV, also shown calculation for ``set 3'' from \cite{Kolevatov:2012vu} with $C_1=C_2=0.5$, $R_1=R_2$. }
\end{figure}

\begin{figure}[!h]
 %\centering 
 \hspace{-6mm}\hbox{\includegraphics[width=0.4\hsize]{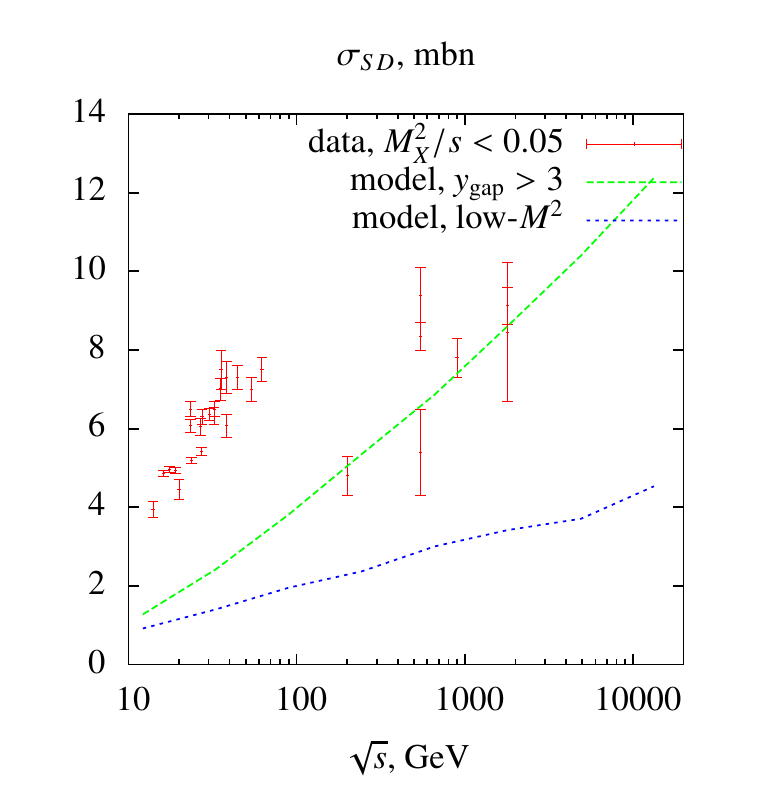}\hspace{-8mm}\includegraphics[width=0.4\hsize]{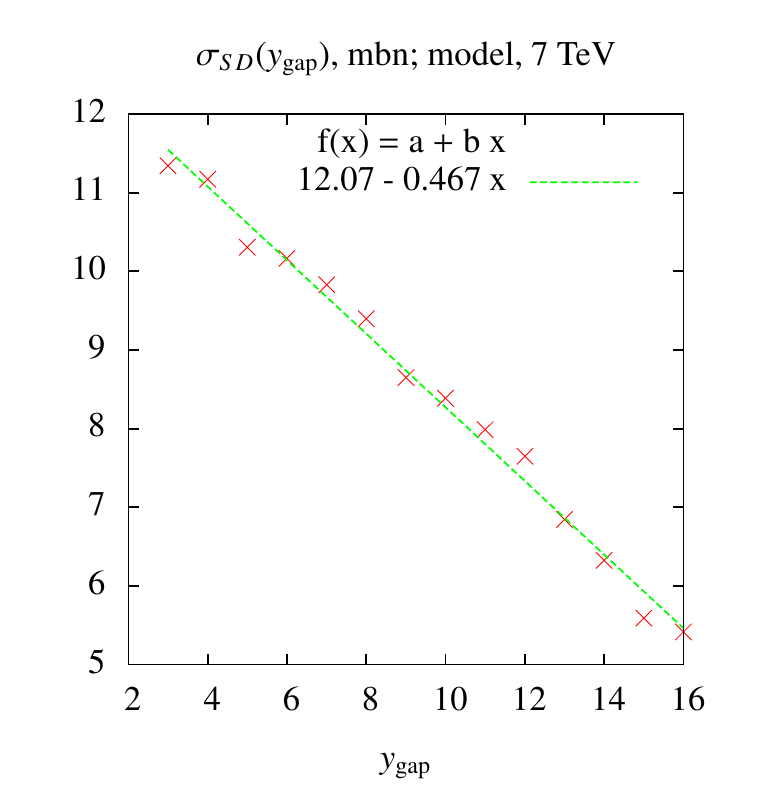}\hspace{-10mm}\includegraphics[width=0.4\hsize]{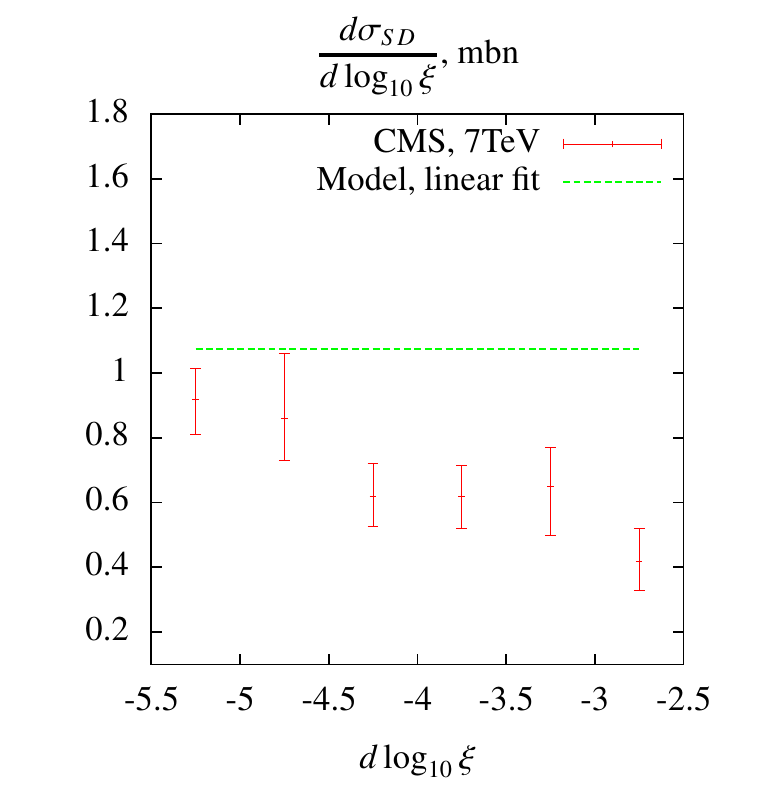}}
 \caption{Model results for the single-diffractive cross section.}
\end{figure}

In fig.~3 (left plot) we show results for single diffractive cross section. Including contributions with interaction of secondary Reggeons (of the type $PPR$, $RRR$ etc.) vanishing at high $\sqrt s$ could improve the situation with fitting the data at low c.m. energies which are at the moment poorly described by the calculation. Another thing is that currently the value of $r_{3P}$ is taken from \cite{Kaidalov:1979jz} where it was extracted from low energy data in the triple Pomeron approximation without account of multipomeron exhanges and loops. Considering $r_{3P}$ as another fitted parameter and including single diffraction data into the fit could also improve the description of inelastic difraction cross section.

The single diffractive cross section as a function of the rapidity gap (fig.~3, center) shows approximately linear behaviour (within the accuracy of the numerical calculation). As $y_{\rm gap} = - \ln \xi$ with $\xi=M_X^2/s$ and $d/dy_{\rm gap} = -M^2_X d/dM_X^2$, this is consistent with $1/M_X^2$ scaling of the $d\sigma_{SD}/dM^2_X$ mentioned above. In the right panel of fig.~3 the slope of the linear fit is plotted versus the CMS data on single diffraction \cite{CMS:2013mda}. At present stage one can speak only about the qualitative agreement of the all-loop calculation results with the diffractive cross sections data.

\section*{Acknowledgement}
Authors are deeply thankful to Oleg Kancheli for the numerous illuminating discussions and helpful advice. RK also thanks the laboratory SUBATECH and Gin\'es Martinez personally for covering travel expences related to visiting the workshop. The work of RK was supported by ``Agence Nationale de la Recherche'', grant ANR-PARTONPROP.

\end{document}